\documentclass[12pt,preprint]{aastex}

\shorttitle{Point sources and CMB experiments}
\shortauthors{E. Pierpaoli }
\begin{document}

\title{Point sources in  the MAP sky maps }

\author{Elena Pierpaoli \altaffilmark{1}}
\altaffiltext{1}{Physics Department and Astronomy Department, Princeton University, Princeton, NJ, 08544}

\begin{abstract}
We discuss  point sources  foregrounds for the MAP experiment.
We consider several  possible strategies for removing them  and 
we assess   how the statistics of the CMB signal are affected by the 
residual sources.
Assuming a power law distribution for the point sources, we 
propose a method aimed to determine the slope of 
the distribution from the analysis of the moments of the observed maps.
The same method allows for a determination of the underlying CMB variance.
We conclude that the best strategy for 
point sources finding is  the simultaneous thresholding
of the filtered map at all frequencies, with a relatively low threshold. 
With this strategy, we expect to find 70 (95)\% of the sources 
above 3 (4) $\sigma$.
Assuming the most conservative case for point sources detection, 
the recovered slope of the point sources distribution 
 is $2.55 \pm 0.15$, for a fiducial $n=2.5$ value.
The recovered CMB pus noise map variance is within $0.2\%$  from the real one,
with a standard deviation of $0.31 \%$, while 
Cosmic variance contributes $2.2\%$ to the same CMB plus noise map.

\end{abstract}

\keywords{cosmic microwave background: extra galactic foregrounds }

\section{Introduction}

The upcoming cosmic microwave background (CMB) experiments
 MAP (Bennet et al., 2003) and Planck (Mandolesi et al., 2002) face the ambitious task of determining 
several cosmological parameter with a high level of accuracy.
Their ambitious goals require  the ability in cleaning the maps 
from the galactic and extra-galactic foregrounds (Refregier and Spergel, 
2000). 
In this paper we address the issue of the possible performance the satellite 
MAP can have  in subtracting a specific extra-galactic foreground, that is 
the point sources. 

Point sources cannot be avoided 
by sky cuts, and are the main source of uncertainty at small angular 
scale where, even when all bright sources are identified and subtracted, 
their  residual contribution may dominate over Cosmic variance.
For example, point sources have been the main foreground issue for the 
CBI experiment (Myers et al.,2002), where they have been the only foreground 
explicitly modeled and considered in the overall likelihood.

Two approaches are commonly adopted for point sources identification:
{\it ii)} the use of exiting point sources catalogs; {\it i)} the 
 direct
search on the data map or on a filtered version of it.
In this paper we will discuss the latter.

The simplest version of the direct search consists in straight thresholding of
 the data. This naive approach only finds very bright sources. An improved 
 version consists in first filtering the data with an appropriate filter 
that maximizes the 
signal of the point source peak with respect to the {\it noise}
 intended as CMB signal plus instrumental noise (Tegmark and de 
Olivera--Costa 1998). 
Here the knowledge of the CMB and the instrumental power spectra are assumed 
(see also Chiang et al. 2002 for a diferent approach). 
Alternatively, it is possible to look for point sources in a map
obtained by subtraction of two given frequency maps.
The thresholding procedure selects from these maps a subsample of pixels 
that are more likely to be contaminated by point sources.
This subsample may be used as the point sources catalog (if the threshold
is high enough to avoid fake detections) or as a useful template
for further investigations with alternative techniques.
This subsample is particularly useful for point sources detection 
in  all sky experiments 
like  MAP and Planck, whose maps contain a large number of pixels.
In this paper we   discuss  the performances of the 
 thresholding  technique
when applied to an experiment with the  characteristics of MAP.
More specifically, for each method described above
we  discuss which fraction of the sources are
selected after thresholding the frequency maps at different levels and
how many pixel in the map are selected by the same procedure. 
We suggest a method aimed to maximize the former and minimize the latter.

Alternative techniques have been proposed in the literature. For example,
 Naselsky et al. (2002) proposed 
a generalization of the filtering method to multi-frequency experiments,
 claiming  an increased signal to noise ratio when filtering an appropriate
linear combination of the observed maps. 
A possible limitation of this method is the needed assumption on the point
source frequency behavior.

Maxican--hat wavelets have also been applied in subtracting the point sources 
(Cayon et al. 2000, Vielva et al. 2001a),
 alone or in conjunction with maximum--entropy (Vielva et al. 2001b).
This technique applies to single frequency maps, and performs similarly to
the filtering method. 
In fact, the filters used in the two methods, although derived
with different prescriptions, end up to be similar.

We  then address the issue of the statistical 
contamination of the residual sources in the CMB map, estimating the accuracy
with which the CMB variance can be recovered if all sources above a certain
threshold are removed from the observed  map.
We then briefly discuss the impact of residual point sources on the power spectrum.

We show how the moments of the observed map  point distribution function
can be used to infer characteristics of the point sources population,
namely the slope of the point sources flux distribution.

The paper is organized as follows:
in section \ref{sec:simu} we present the characteristics of the simulation;
in section \ref{sec:resu} we present the results obtained, that is the 
efficiency in finding point sources (\ref{subse:ptfin}), the impact of the
residual signal  on the statistics of  the CMB map and power spectrum 
and the  performance in  the determination  of the slope of the 
point sources distribution  (\ref{subse:resi}, \ref{subse:powsp}).
Section \ref{sec:con} is dedicated to the conclusions.

\section{Simulations} \label{sec:simu}

We simulate CMB sky  maps  with the characteristics 
of the MAP  experiment (see tab.~\ref{tab:mapc}).
For this purpose we use a cold dark matter power spectrum,
 and  uniform Gaussian noise.

\begin{table}
\begin{center}
\caption{ The characteristics of the MAP experiment: column one is the 
frequency of observation, column two is the FWHM of the beam in degrees, 
column three (four) is the expected noise standard deviation for 2 years (six months) observation and for a $0.3 \times 0.3$ degrees pixel in  $\mu K$. \label{tab:mapc}}
\begin{tabular}{crrr}

\tableline\tableline
$\nu ~(GHz)$ & $\theta_{FWHM}$  &  $\sigma_{n}~(\mu K, 2 yrs) $ &$\sigma_{n}~(\mu K, 6 mo.) $  \\
\tableline
22& 0.89  & 35~~~~~ & 70~~~~~ \\
30 &  0.646&    35~~~~~ & 70~~~~~ \\
40 &  0.504  &  35~~~~~ & 70~~~~~ \\
60 &   0.312 &  35~~~~~ & 70~~~~~ \\
90 &     0.214  &  35~~~~~ & 70~~~~~ \\
\tableline
\end{tabular}
\end{center}
\end{table}

Simulations were performed in the Healpix pixellization scheme 
\cite{Heal}, with resolution N=256. The typical pixel side is 0.2 degrees, 
 which is comparable to the highest resolution of the experiment.

As for the point sources, we consider the following  flux dependence:
\begin{equation}
d {\cal N}(S)= A~S^{-n}~dS
\label{eq:neq}
\end{equation}
with $n=2.5$ (expected values of $n$ range from $1.5$  to $2.5$). 
We consider two  different normalizations,
 which amount to have  178 or 356  sources respectively above 
 $1$ Jy at $90$ GHz. These normalizations correspond approximately with 
the numbers proposed by  Holdaway et al. (1994) and Toffolatti et al. (1998)
 respectively. 
We present most  results are presented in the highest normalization case, 
while we mainly use the lower  for testing.

At low fluxes the power law in eq.~\ref{eq:neq} is expected to break down,
the exact flux at which this happens and the shape of the distribution
at lower fluxes is unknown.
We take this into account by  simulating only sources with flux above 
$S_{min}=0.01$ Jy at $90$ GHz.
We then extrapolate 
the $90$ GHz simulated  sources to lower frequencies
assuming an approximately flat spectrum:
\begin{equation}
S(\nu) \propto \nu^{-\gamma}
\end{equation}
with $\gamma$ randomly chosen in the range $[-0.3 , 0.3]$.

We convolve the point sources maps  with the appropriate beam before
summing it to the CMB plus noise (CMBn) ones.
We do not introduce other foregrounds in our modeling, therefore  the
 estimates we make should be 
considered to hold for regions that have a negligible contamination of other 
foregrounds.

We  have about $5540$ sources in our catalog with the lower normalization, 
and the double with the higher. In table \ref{tab:tab2} 
we show how their flux compares to the standard deviation of the CMBn map
at $90$ GHz.

\begin{table}
\begin{center}
\caption{The distribution of simulated sources fluxes at $90$ GHz with respect
to the standard deviation in the CMB plus noise (CMBn) map. The first column is the cut in units of $\sigma_{CMBn}$, the second and third columns are respectively   the 
number and the fraction of simulated sources within that range. These numbers 
correspond to the higher normalization of the point sources distribution.
\label{tab:tab2}}
\begin{tabular}{crrr}
\tableline\tableline
$\sigma_{CMBn}$ threshold & $N_{s}$ & $F_{s}$ \\
\tableline
$< 1$ &      7548 &     0.674 \\
1 -- 2 &   2336 &   0.209 \\
2 -- 3 &    617 &    0.055 \\
3  -- 4 &     238 &    0.021 \\
4 -- 5 &     135 &    0.012 \\
5 --  6 &      69 &  0.006 \\
6 -- 7 &     48 &   0.004 \\
7 -- 8 &      40 &   0.003 \\
8 -- 9 &      21 &  0.0019 \\
9 -- 10 &     25 &   0.002 \\
10 -- 11 &      13 &   0.001\\
11 -- 12 &      9 & 0.001\\
$ > 12$  &    92 &  0.008\\
\tableline

\end{tabular}
\end{center}
\end{table}

\section {Results} \label{sec:resu}
In this section we present our results. In sec.~(\ref{subse:ptfin}) we analyze
and compare different methods for point sources finding. In 
sec.~(\ref{subse:resi}) we use 
the statistics of the observed map
in order to assess
the underlying CMBn map variance and the 
 characteristics of the point sources distribution.
Finally, in  sec.~(\ref{subse:powsp}) 
we show the impact of residual point sources
on the power spectrum.

\subsection{Point sources identification}{\label{subse:ptfin}}

In this section we compare three techniques for point sources identification. 
Namely, we perform a thresholding on the following maps:
{\it i}the raw data (method I);
{\it ii}  the maps obtained after applying the filter introduced by 
Tegmark et al. (1998) (method II);  a map 
obtained by the subtraction of two maps at different frequencies (method III).

More specifically, in order to apply method II
we  convolve the observed image in each given frequency
with a window function whose  spherical harmonics coefficients are:
\begin{equation}
W(l)= {{b(l)} \over {b(l)^2 C(l) + C_{noise}(l)}}
\label{eq:wl}
\end{equation}
where $b(l)$ is the spherical harmonics of the beam and $C(l)$ and 
$C_{noise}(l)$ are the power spectra of the CMB and noise respectively.
This filtering has the advantage of subtracting the background and 
lowering the noise. As a result, it reduces the bias one may have in 
thresholding  the data directly, where sources of the same intensity are
less (more) likely to be detected if they sit in a CMB valley (hill).

In order to apply method III we deconvolve   the $90$ GHz map, and then 
we subtract it from each other frequency map  after convolving it with the 
beam of the experiment at that frequency.
This  strategy is aimed to subtract the CMB signal. The potential problem
is the increased noise in the output maps.

After filtering the data or combining the observed maps as explained above,
point sources candidates are selected by thresholding the output maps 
and retaining all pixels above a given threshold.
A  candidate  pixel can be confirmed as a  point source center
 by looking at the radial profile of the
map around it, or at the frequency dependence of that particular pixel in
all frequency maps.
This procedure is necessary when dealing with real data, and it allows to 
produce a catalog of objects with a certain likelihood of being a point 
source. The detection probability will ultimately depend upon the likelihood
method applied and the modeling of the source considered.
Clearly, point sources that didn't make it above the threshold are certainly
not going to be detected. While lowering the threshold allows for more 
sources to get into the selected set of pixel, it would also allow for
a greater number of pixels not associated with point sources to enter
the selected sample.
It is therefore important to decide which method
retains the greatest number of sources locations among the selected pixels
and compare it to the total number of pixels selected at any given threshold.
By minimizing the number of pixel selected while maximizing the one containing 
the point sources centers we have  tentative point sources catalog
that can be used for efficient point sources confirmation with alternative 
methods.

We proceed  by comparing method I and II first.
We filtered each frequency map independently with the filter in 
eq.~\ref{eq:wl}, and performed
a thresholding
the filtered map and of the raw data at different levels 
(parametrized by the standard deviation
of the map in hand $\sigma_{thr}$). 
We then counted the point sources found above a given level \footnote{
We consider a source to be found if the pixel corresponding to its center
has an intensity above the threshold}, and how many pixels are in the 
subsample.

\begin{figure}
%\plotone{effic.ps}
\plotone{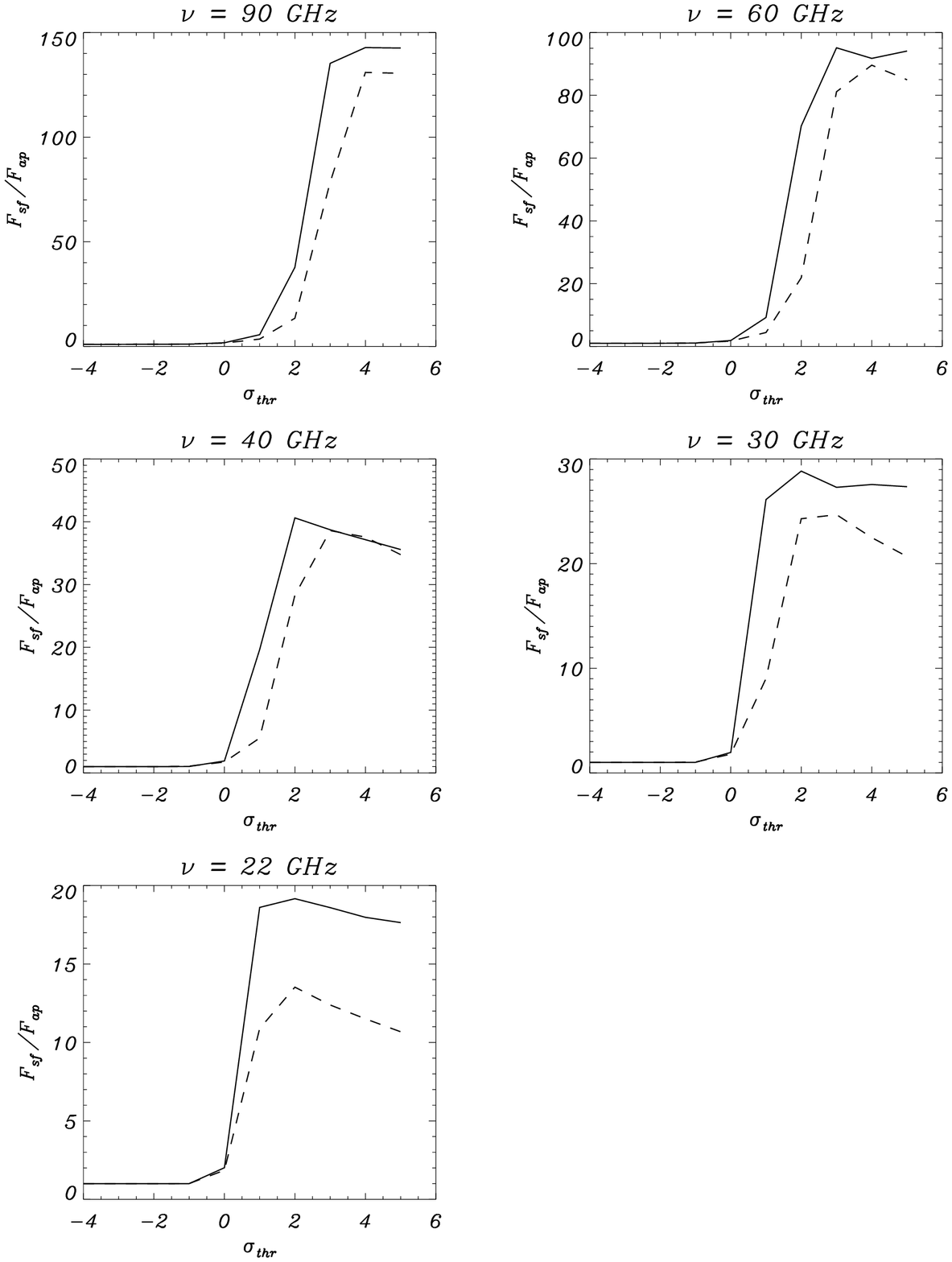}
\caption{Efficiency of the thresholding procedure. The y axes shows the
 fraction of point sources found over the fraction of pixels in the map above
 threshold $\sigma_{thr}$. The solid line corresponds to the  filtered map 
(method II) while the dashed to the observed data map. 
$\sigma_{thr}$ is the standard deviation of the map under analysis. 
\label{fig:effic}}
\end{figure}

\begin{figure}
%\plotone{wie_frac.ps}
\plotone{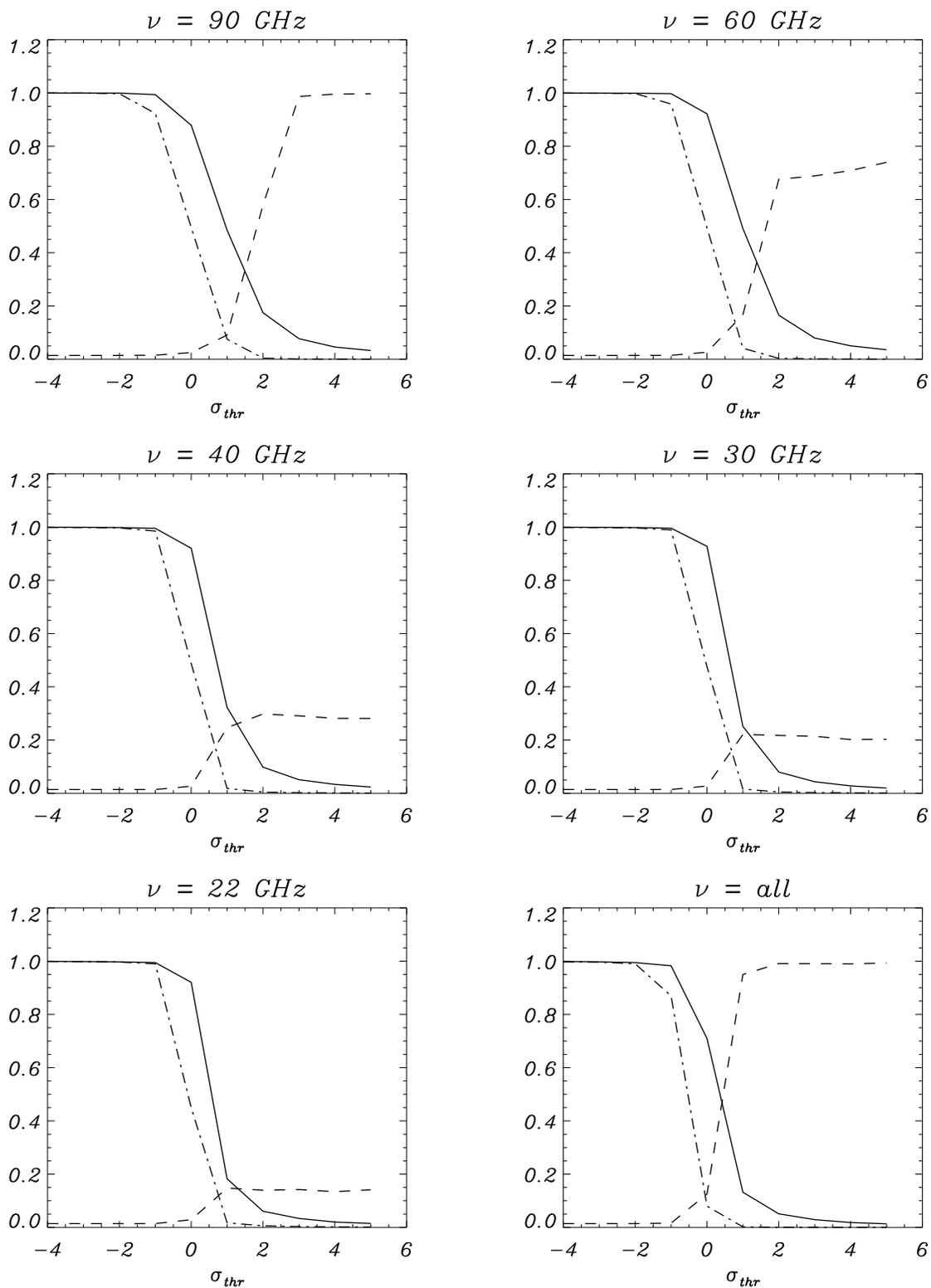}
\caption{Performance of  the filtering  method  (method II).
Various fractions are plot as a function of the threshold applied (in
units of  the filtered map standard deviation).  
  Solid line: fraction of sources found. Dashed line: number of pixel 
containing point sources over number pixel considered. 
Dot--dashed line: fraction of pixel considered over total number of pixel.\label{fig:wiefrac}}
\end{figure}

In figs.~\ref{fig:effic} and~\ref{fig:wiefrac} we show the 
performance of the two methods in finding 
the point sources. The plot in fig.~\ref{fig:effic}
 reports the ratio between the fraction
of point sources found $F_{sf}$ with respect to the fraction of pixel selected
$F_{ap}$ as a function of the threshold $\sigma_{thr}$.
The frequency where source finding is more efficient is $90$ GHz.
We note that in the filtered map there is a big increase in efficiency  
for $\sigma_{thr}$ between 1 and 3 (this value  is shifted between 2 and 4
for the data map).
Looking at  fig.~\ref{fig:wiefrac} it is clear that
the threshold  on the $90$ GHz map is set at
$\sigma_{thr} = 2$ only about $20\%$ of the total sources can be found.

\begin{figure}
%\plotone{sfrac90GHz.ps}
\plotone{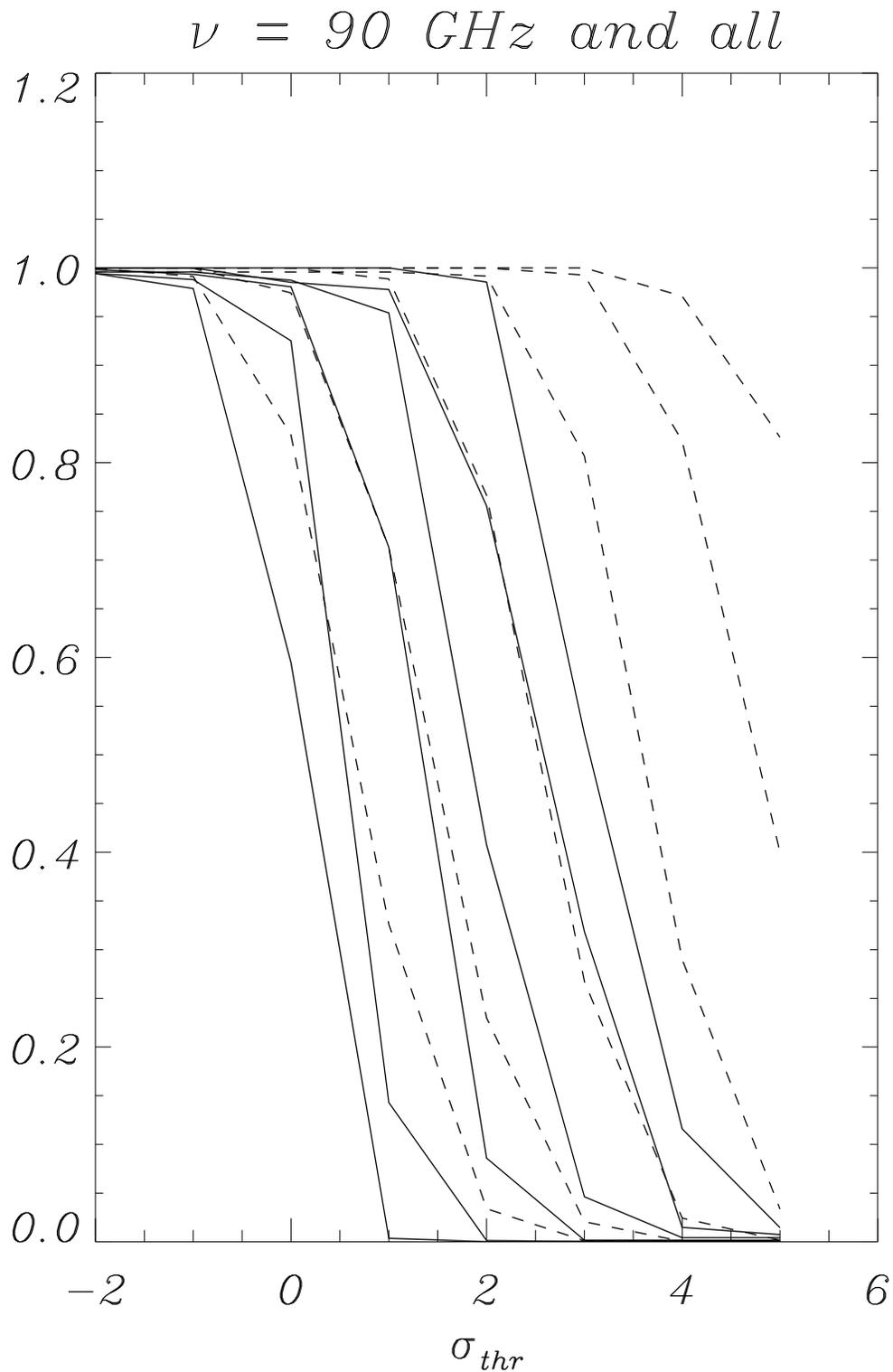}
\caption{Fraction of sources found in the filtered 90 GHz map (dashed line)
and in all maps (solid line) as a function
of $\sigma_{thr}$ (standard deviation of the filtered map). From left to right,the lines correspond to sources below 1 $\sigma_{CMBn}$, 
between 1 and 2, 2 and 3, 3 and 4, 4 and 5, 5 and 6. 
\label{fig:sfrac}}
\end{figure}

Lowering the threshold at 1 $\sigma_{thr}$ allows to detect about $1/2$ of
 the sources, including all sources above $3 \sigma$ 
(see fig.~\ref{fig:sfrac}) and $85\%$ of sources above  $2 \sigma$.
However, one would still be left with about $10\%$ of the total number 
of pixel to analyze (fig.~\ref{fig:wiefrac}, upper left panel). 
This problem can be avoided by selecting the pixels according to whether 
or not they are above a certain threshold at all frequencies.
The bottom right  panel in fig.~\ref{fig:wiefrac} shows that if one considers
pixels with signal above $0.5 - 1 \sigma_{thr} $  at all frequencies, one may
still find $20 - 50 \%$ of all sources with less pixels to analyze. 
The ratio between the number of pixels occupied by point sources centers and 
the number of pixels above the threshold ($R_{occ}$; dashed line in 
fig. ~\ref{fig:wiefrac}) ranges between 
0.5 and 0.95, while in order to get a similar amount of sources from the 
90 GHz map alone that ratio  is 0.05 -- 0.2.

In the same thresholding range, if we considered  all the observed maps  
instead of the filtered ones we would still  be able to find  
0.2 -- 0.4 of the total number of sources, but with a much noisier signal:
the  ratio $R_{occ}$ would be 0.15 -- 0.20.

In fig.~\ref{fig:sfrac} we report the fraction of sources found 
above a certain threshold  in all filtered maps  as a function of source 
intensity.
If $\sigma_{thr}$ is chosen to be one, $95 \%$ of sources above $4 \sigma$
and $70 \%$ above $3 \sigma$ can be found. 
Due to the steepness of the curves, lowering the threshold to $0.5 \sigma$
 would imply the possible detection of $90 \%$ of sources above $3 \sigma$ and 
$65 \%$  above $2 \sigma$.
When dealing with the real map, however, there is the additional complication
of confirming the point sources by looking at their profile in the map.
This procedure is likely to be more difficult for weaker sources.
The number estimated here are therefore to be considered optimistic for 
weak sources.

When analyzing the maps obtained as differences of two frequency maps (method III), we find different results according to the frequency considered.
The efficiency of point sources finding 
on the $(60-90)$ GHz map is very low, the fraction 
of point sources found being 
equal to the fraction of pixels analyzed for any threshold.
When maps at lower frequencies are considered, the performance of this method
is similar to the filtering one (method II).
 However, the combined information from 
different frequencies is lower because in practice this procedure gives up
two frequency maps. 

We conclude that the  best  strategy consists in using
  the combined information from the filtering method (method II), with a 
threshold 0.5 -- 1 $\sigma_{thr}$.

Note that the  procedure proposed  does not assume any particular 
frequency dependence 
for the point sources. It just exploits the fact that point sources are 
expected to be in the same locations in the different frequency maps,
 while the peaks in the noise distribution are uncorrelated
between frequency maps.

The present analysis was based on some assumptions implicit in the simulations,
like the pixel size or the frequency dependence of the point sources.
As for the pixel size, we assume the biggest possible considering the 
highest experiment resolution.
If we chose to perform a higher resolution simulation the noise variance
would increase, while the signal from point sources would 
remain the same.
Since method II performs a convolution of be map with a window function
that has a characteristic smoothing scale of the order of the beam size 
at that frequency, using a finer resolution wouldn't change the performance.
In methods  I and II there would be an increase of spurious detections at 
any given threshold due to the higher pixel noise, resulting in a less 
efficient  point sources finding.
Given our conclusion on the preferred method, we don't investigate this issue any further.
 
As for the frequency dependence of point sources, the approximately
 flat spectra sources  we simulated seem to fit the majority of radio sources
(Sokasian et al. 2001). 
The Rayleigh--Jeans temperature fluctuation produced by point sources scales as
$\Delta T =  S(\nu) \Theta_{FWHM}^{-2} \nu^{-2}$,  and since for MAP we have
$\Theta_{FWHM} \propto \nu^{-1}$ 
there is no obvious frequency  map where to look for flat spectrum sources. 
Should future observations suggest that a considerable fraction of sources 
show significant departures from this frequency dependence, 
a different strategy 
might be needed. In particular, if the spectra show a bump in the frequency
range of interest, there might be a preferred frequency 
map, different from the 90 GHz one,  for point sources detection.
The method presented here is however not dependent on the point sources
 spectrum, and can in principle be investigated with different simulations.

\subsection{Using moments to determine residual point sources contribution} {\label{subse:resi}}
After having identified and subtracted the bright sources, the remaining ones
still affect the real space statistics  of the CMB that we want to measure.
We shall use this dependence in order to infer both  properties of the 
point sources distribution and the expected variance of the CMBn point
distribution function.

Having assumed Gaussian fluctuations, we expect the point distribution
function of the  CMB signal to be 
completely described by its variance. For a full sky experiment we have:
\begin{equation}
\sigma^2_{CMB}=\sum_l {{2~l+1} \over {4~\pi}} C(l)b^2(l).
\label{eq:sig2cmb}
\end{equation}

While an unbiased estimator for this quantity exists,
Cosmic variance sets a limit to the precision  with which the variance is
determined: 
\begin{equation}
var(\sigma^2_{CMB})= \sum_l {{2~(2~l+1)} \over {(4~\pi)^2}} C^2(l)b^4(l).
\end{equation}
In the specific CDM case we are adopting here and for the beam of the 
90 GHz map, we have:
\begin{equation}
\sqrt{(var(\sigma^2_{CMB}))}/\sigma^2_{CMB}=3.0 \%.
\end{equation}
Given the additional contribution of the noise (for our pixellization and 
2 years of data), 
the Cosmic variance is expected to be $2.2 \%$ of the CMB plus noise signal
($\sigma^2_{CMBn}=\sigma^2_{CMB}+\sigma^2_{noise}$).

Faint point sources contaminate the map rendering the CMB+noise
 statistics 
non--Gaussian.
Estimators of non--Gaussianity may rely on asymmetries of the point distribution
function and the study of its moments (Rubin--Martin \& Sunyaev, 2002),  
or on the study of the bispectrum and trispectrum in Fourier space 
(Komatsu et al 2001).

In this work we concentrate on 
 the first four moments of the point  distribution function of the map after
sources subtraction.
On the $90$ GHz map we performed a progressive subtraction of point 
sources  above a certain threshold $\sigma_{cut}$ (expressed in units of 
$\sigma_{CMBn}$).
All pixels in a ring of radius of $1$ FWHM from the point source location 
were removed from the map  in order to avoid residual signal from 
the convolution process.
  
\begin{figure}
\plotone{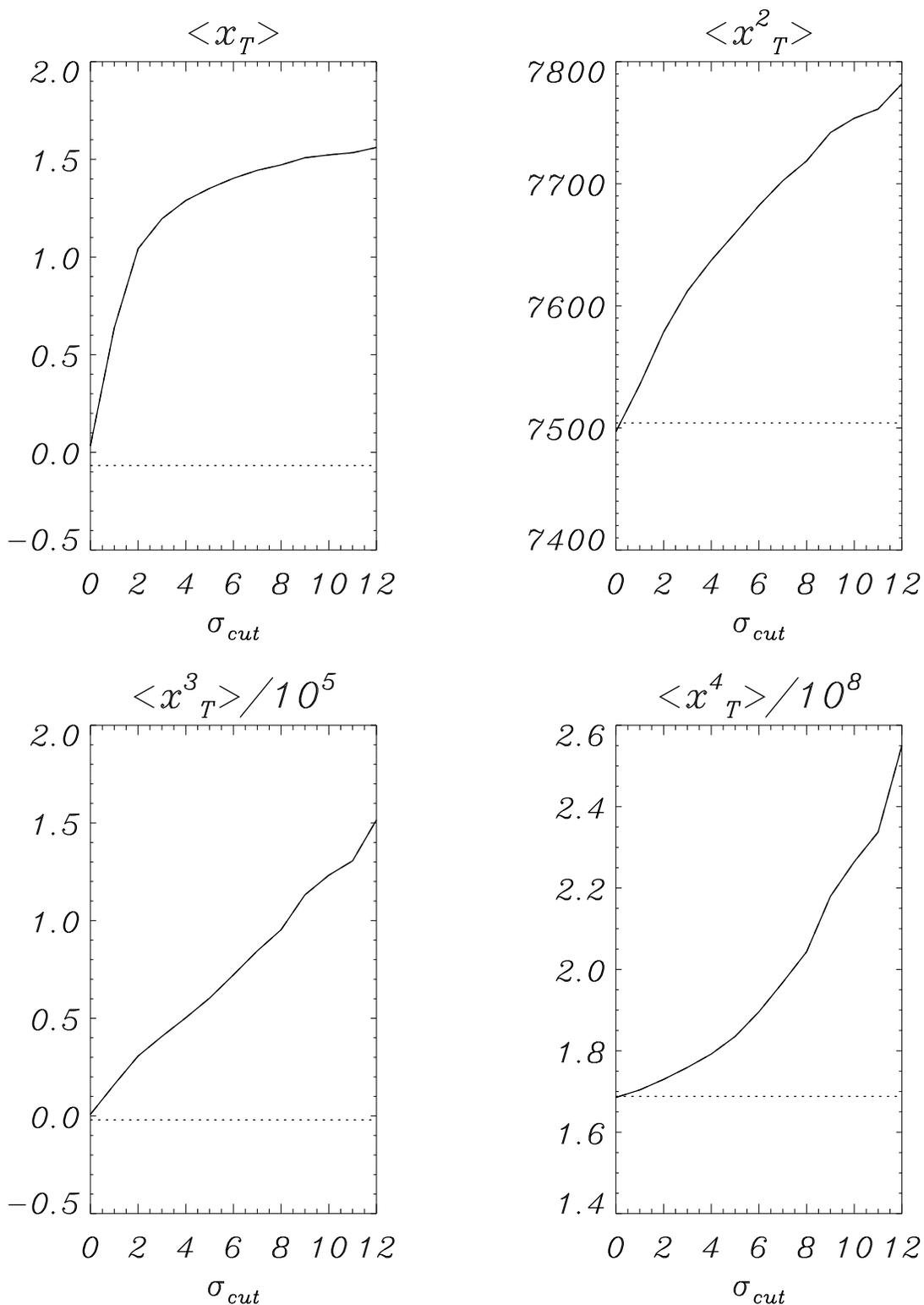}
\caption{The first four moments  of the 90 GHz observed map once point sources above $\sigma_{cut}$ have been subtracted. The dotted  and solid lines are the moments of the CMBn and observed maps respectively. The map
is in Raleigh--Jeans temperature, the y axis units vary accordingly.
 \label{fig:mean}}
\end{figure}

In figs.~\ref{fig:mean}  we show, for a particular map,
 how the various moments  are affected by the remaining point sources.
Although bright sources are rare, their impact on the statistics of the 
map is most relevant. 
For example, in the realistic case that all sources above 
$4 \sigma_{CMBn}$  are removed, the estimated variance 
of the cleaned map would be within $2\%$ of  $\sigma_{CMBn}^2$.
This means that residual point sources potentially affect the final
distribution inducing an error on the variance comparable to the Cosmic 
variance.

The other relevant feature of fig.~\ref{fig:mean} is  the shape 
of the curves, which is linked to the characteristics of the point sources 
distribution.
In the following we will discuss how how well 
it is possible to determine such distribution and the variance
of the CMB by looking at the moments of the maps.
The moments plotted here are non--centered, that is they are calculated as:
$\langle x^\beta \rangle=\int{x^\beta~{\cal N}~dx}$, where ${\cal N}$ is the normalized 
pixel distribution. 
For a Gaussian field with zero mean, as the simulated CMBn map is,
we would expect $\langle x \rangle=0$, $\langle x^2 \rangle=\sigma^2$, 
$\langle x^3 \rangle=0$, $\langle x^4 \rangle=3 \sigma^4$.
In reality, the CMBn map itself deviates from this formulae because of Cosmic
variance, so that
the Gaussianity hypothesis cannot be fully exploited.
In the following we assumed  $ \langle x \rangle=0$ and 
$ \langle x^4 \rangle=3 \langle x^2 \rangle$, but put no 
constraints on $\langle x^3 \rangle$.
The final map is a sum of the CMBn map ($x$)  and the point sources ($e$):
 $x_T=x+e$. Given the assumptions above, the moments of the total map 
can be expressed in terms of moments 
of $x$ and $e$ as follows:
\begin{eqnarray}
& &\langle x_T \rangle = \langle e \rangle\\
& &\langle x_T^2 \rangle =\langle x^2 \rangle +\langle e^2\rangle\\
& &\langle x_T^3\rangle =\langle x^3\rangle +\langle e^3\rangle +3 \langle x^2
\rangle
\langle e \rangle\\
& &\langle x_T\rangle =3\langle x^2\rangle (\langle x^2\rangle +2\langle e^2 
\rangle)+\langle e^4 \rangle\\
\label{eq:intmom}
\end{eqnarray}

where $e$ is a function of the intensity of the sources removed: 
$e(\sigma_{cut})=\int_{S_{min}}^{S_{cut}}{{{\cal N}_{pts}}~dx}$, with $S_{cut}=
\sigma_{cut}~\sigma_{CMBn}$.

Any given moment is therefore a function of the slope $n$, the normalization
$A$ and the minimum flux of the sources $S_{min}$, none of which is supposed to be known a priori.
We performed 100 simulations of point sources map and added them to 
a noisy CMB one.
For any given simulated map, we calculated the first four moments as 
a function of $\sigma_{cut}$. We then fixed $n$ 
and performed a simultaneous fit to
all moments using  the formulas in eq.~(\ref{eq:intmom}).
As a conservative approach,  we considered to be able to detect only sources
above $4 \sigma_{CMBn}$, and therefore considered to know the curves in fig.~
\ref{fig:mean} only for $\sigma_{cut} \ge 4$.
The parameters of the fit were: $\langle x^2 \rangle$,$\langle x^3 \rangle$,
 $A$ and $S_{min}$, and the 
likelihood was simply the sum over all moments and all points of 
the percentage difference of the fit to the data.
We then selected the $n$ value which minimized the likelihood, 
and its associated parameters at the best fit.

In figs.~(\ref{fig:enne}) and  (\ref{fig:histo}) and  we show the histograms 
for the $n$ values found and for the assessed $\sigma_{CMBn}$.
As for the $n$ value, the estimated mean and standard deviation are:
 $n=2.55 \pm 0.15$.
The significant standard deviation
 of the distribution is due to the fact that  there is  
degeneracy between the parameters, in particular between  $n$ and $A$.
 The best fit theoretical curves 
for a fixed $n$ are indistinguishable, so that one should resort to other 
methods to have a more precise $n$ estimate.
An alternative method, not investigated here, could consist in fitting the
positive tail of the observed distribution, in the region where the CMBn
contribution is expected to be negligible. 

On the other hand, the shape of the underlying theoretical curves 
is very well constrained. As a result, the estimated value of the 
CMBn map variance
($\sigma_{est}^2$)is very close to the true one:
$(\sigma_{est}^2-\sigma^2_{CMBn})/\sigma^2_{CMBn}=-0.22 \% \pm 0.31 \%$.
Knowing the noise properties of the instrument, we are therefore 
confident that we can estimate the variance of our underlying CMB map 
with a precision that is far below Cosmic variance.
In fig.~(\ref{fig:histo}) we also note that if we actually knew the precise 
true $n$ value, our estimate for the variance would have been
$(\sigma_{est}^2-\sigma^2_{CMBn})/\sigma^2_{CMBn}=0.11 \% \pm 0.15 \%$.

These results are dependent on the fact that all curves are fitted simultaneously and that the second and fourth moment for the CMB follow the Gaussian
hypothesis.
Fitting the variance curve only doesn't allow neither a good determination of 
the spectral index $n$ nor of the CMBn variance.
When imposing that all curves are fitted simultaneously, improved results 
are obtained only if we impose that the extrapolated values for the second
and fourth moments follow what one would expect for a Gaussian distribution.

The value of $\sigma_{CMBn}$ depends on the pixel noise which in turn is a
 function of the pixel size assumed. Given any initial finer pixellization,
it is possible to rebin the map  to this resolution without loss of cosmological
information. The present procedure can then be applied.

\subsection{The impact of point sources on the power spectrum} {\label{subse:powsp}}

\begin{figure}
\plotone{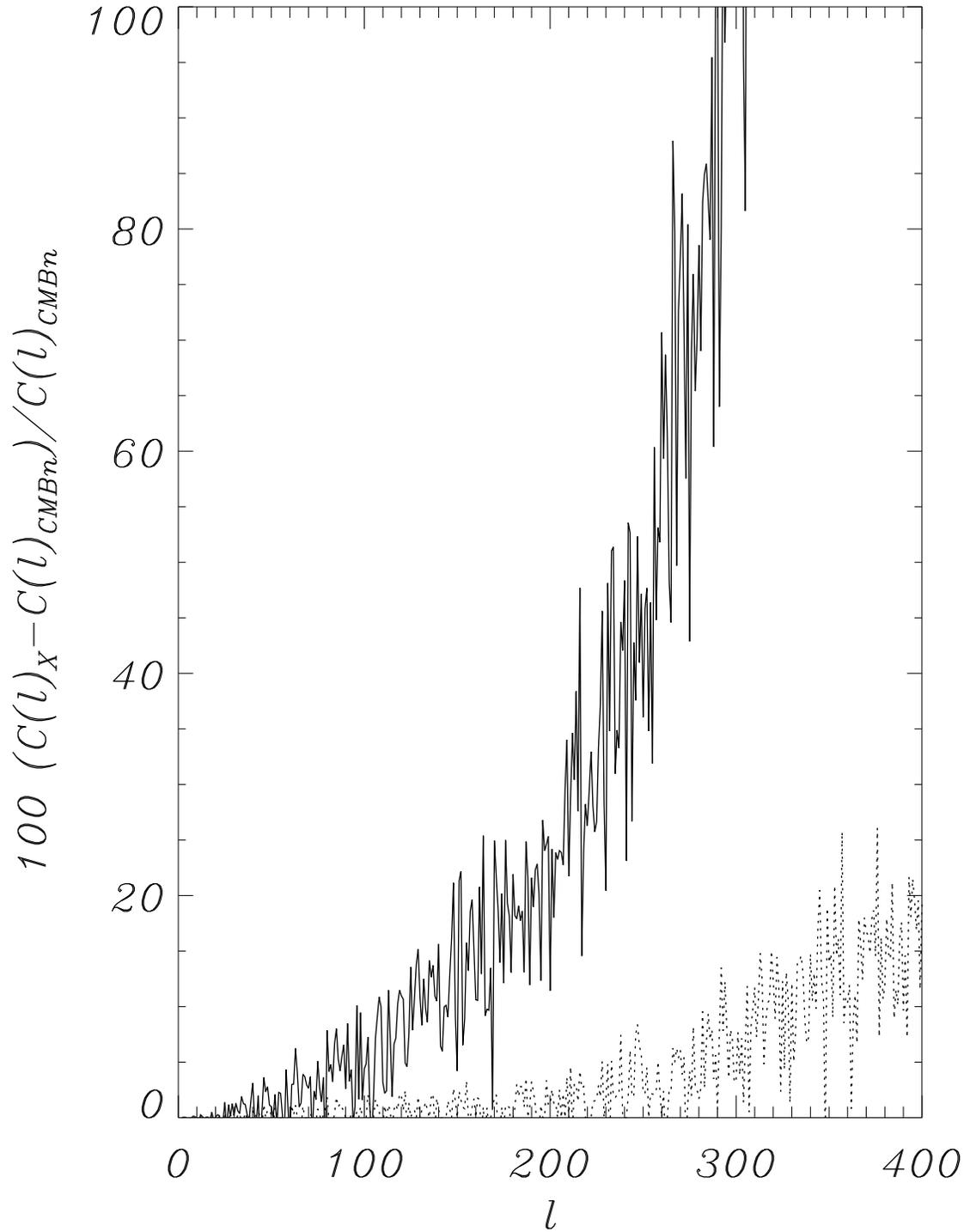}
\caption{
The percentage difference between the power spectrum computed from 
a 90 GHz CMBn map and the corresponding observed one with all 
point sources (solid 
line) and only sources below 4 $\sigma$ (dotted line).
\label{fig:powsp}}
\end{figure}

\begin{figure}
\plotone{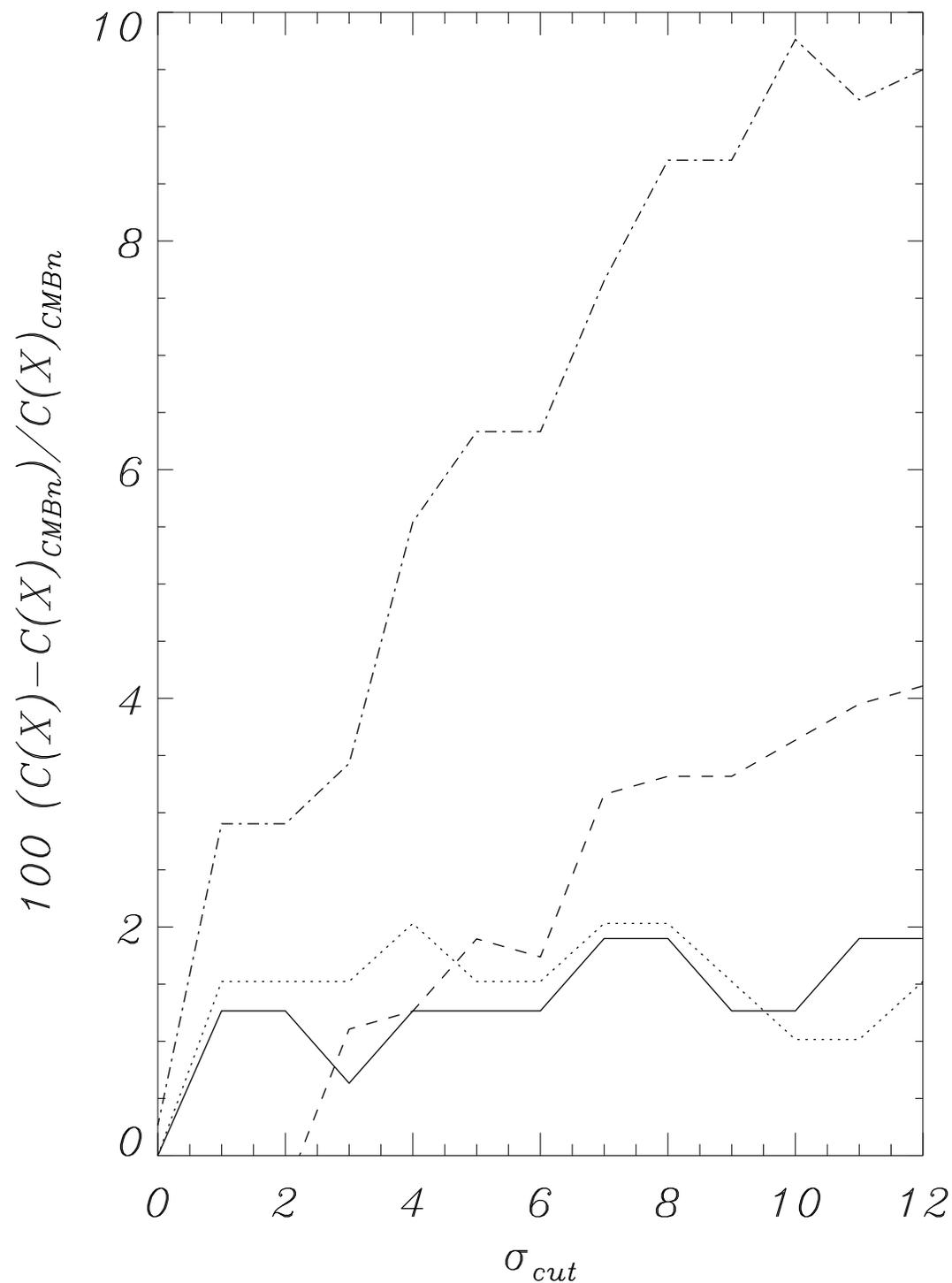}
\caption{
Percentage difference between the power spectra computed from the observed 
90 GHz map (after progressive subtration of point sources) and the underlying 
CMBn map.
The different lines correspond to $l=X=400 ({\rm dot-dashed}), 300 ({\rm dashed}), 200 ({\rm solid}) and 100 ({\rm dotted})$.
\label{fig:powscut}}
\end{figure}

In this section we show the impact of the point sources on the power spectrum,
computed as $C_l=\sum_m{|a_{lm}|^2}$ where $a_{lm}$ are the spherical harmonics of the map. We applied the procedure to the observed map (CMBn plus point sources)
and then to the observed map where sources above a specific threshold have been subtracted.
In fig.(\ref{fig:powsp}) we show, for a particular 90 GHz map, 
the difference between
the power spectrum of the observed map and the underlying CMBn one.
Point sources contaminate the power spectrum at 20\% level at $l \simeq 200$. 
In fig.(\ref{fig:powscut}) we plot, for some particular l's, the percentage difference between the CMBn power spectrum and the one obtained from the maps where
sources above a  threshold  $\sigma_{cut}$ have been subtracted.
Bright sources have the greatest impact on the power spectrum. Subtracting 
sources above 12 $\sigma$, for example, brings the power spectrum from more
than 100\% to only 9\% away 
from the CMBn one at l=400.
 Progressively subtracting fainter sources have a greater 
impact on the higher l's.
At $l \simeq 200$, the impact of point sources is about 2\% quite 
independently from $\sigma_{cut}$ (fig.~\ref{fig:powscut}).
If all sources above 4 $\sigma$ are subtracted from the map,
 the power spectrum is overestimated by about 2\% at $l=200$ and  about 15\% 
at l=400.

\section{Conclusions} \label{sec:con}

We investigated the problem of point sources subtraction from the 
observed MAP multi-frequency maps.

First we analyzed  possible strategies for point sources finding 
without the help of external maps.
We found that the best strategy consists in applying the linear filtering 
in eq.~\ref{eq:wl} to each frequency independently, cutting the maps 
with a relative low threshold ($0.5-1 \sigma_{thr}$) and then considering  
only the pixels that satisfy this constraint in all maps.
This method outperforms similar strategies applied to observed maps and 
to maps obtained by subtraction of two frequencies.
With this strategy we may expect to find  $95 \%$ of sources above $4 \sigma$
and $70 \%$ above $3 \sigma$, as a conservative estimate. 

\begin{figure}
\plotone{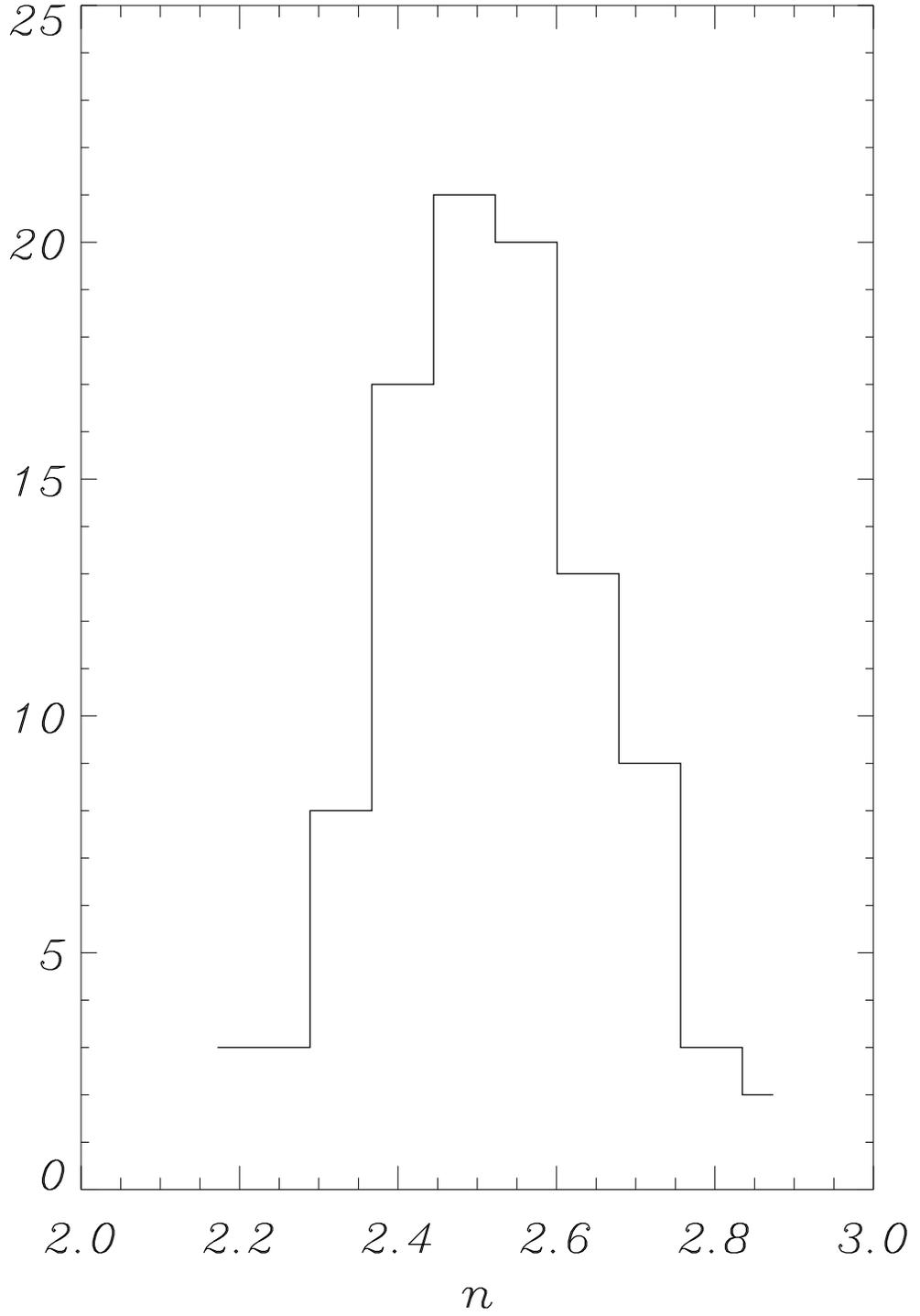}
\caption{
Histogram of the estimates of the slope of point sources $n$ from the 
simultaneous fitting of all moments.
\label{fig:enne}}
\end{figure}

\begin{figure}
%\plotone{histograms.ps}
\plotone{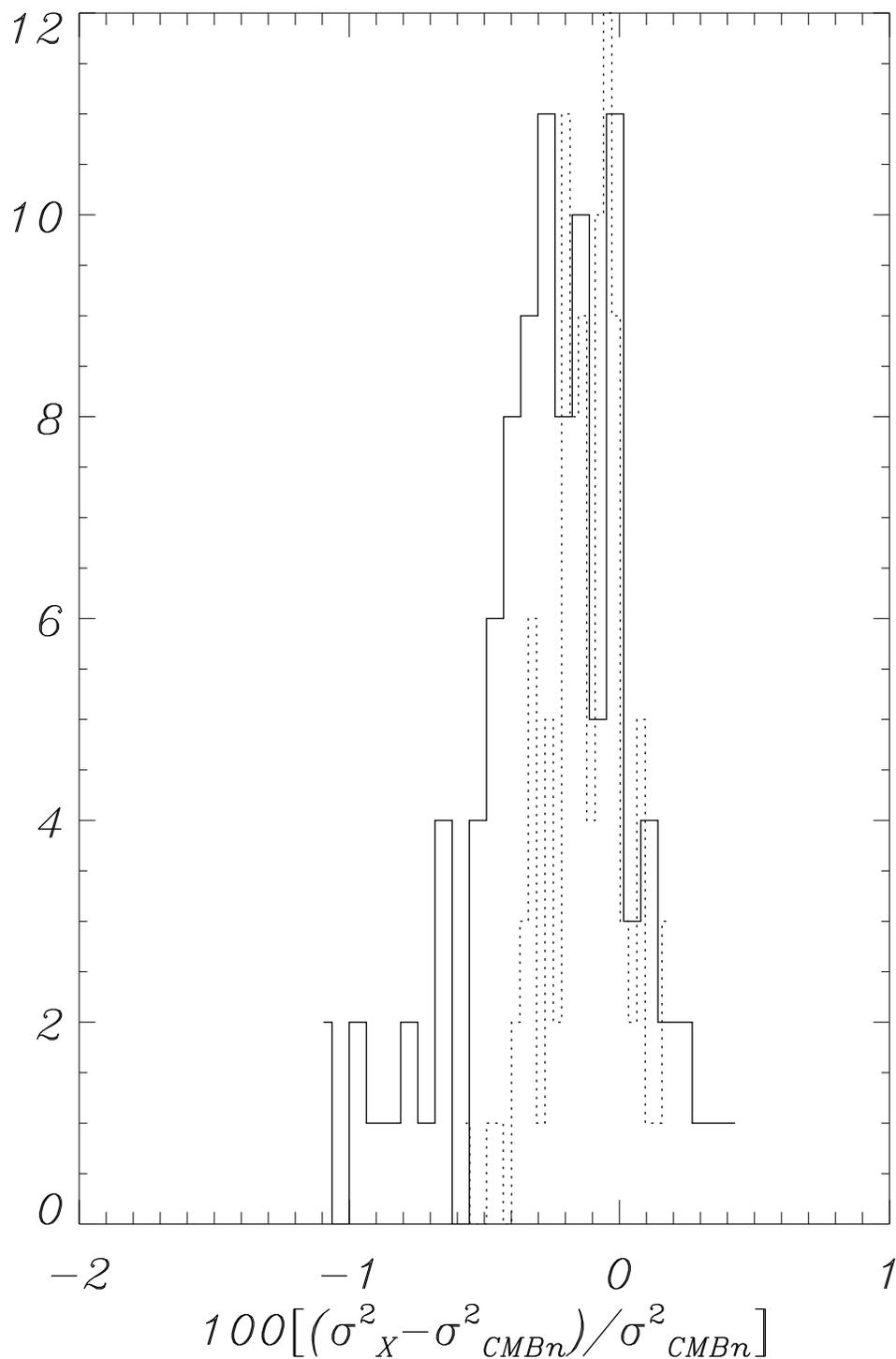}
\caption{ Histograms of the percentage differences between the 
 estimated variance $\sigma^2_X$ and the input map value.
The solid curve refers to best fit values for the $n$ values that minimize
the likelihood (fig.~\ref{fig:enne}). The dotted histogram correspond
to the $\sigma^2_X$ value obtained for a fixed $n=2.5$.
Both histograms are obtained fitting simultaneously all moments 
for $\sigma_{cut} \ge 4$.
\label{fig:histo}}
\end{figure}

Assuming that the brightest sources are going to be found, we addressed 
the issue of how statistics is affected by the residual sources in the map.
 If all sources above $4 \sigma$  are found
the variance of the CMB map is recovered within $2 \%$.
We used the information from the first four moments of the map
to infer the variance of the CMB plus noise map and the slope
of the point sources distribution.
The results are summarized in figs.~\ref{fig:enne} and \ref{fig:histo}.
Extrapolating the variance curve to null source contribution ($0 \sigma$) 
allows to  determine the variance
of the $CMB$ plus noise map within $-0.2 \pm 0.3 \%$. This value should be 
compared with the ratio of  Cosmic variance to 
the CMBn one, which is about $ 2.2 \%$ 
for the CDM  case treated here.
The same strategy allows for a determination of the spectral index 
of the point sources distribution.
For a fiducial $n=2.5$ value, we find 
$n=2.55 \pm 0.15$, where the significant spread is due to 
a  degeneracy with the normalization $A$.

These results are obtained in the  approximation of null contribution
from other foregrounds. While this is likely to be the case close to the 
galactic pole, it may not hold near the plane.
It is always possible to apply this method to submaps of the sky, although 
the lower number of bright sources would probably reduce the performance.

\acknowledgments

E.P. is supported by NASA grant NAG5-11489 and NAG5-10811.
E.P. thanks David Spergel for useful discussions, the anonymous referee
for comments  and 
 the Aspen Center for Physics for hospitality during the preparation
 of this work.

\clearpage


\clearpage

\clearpage

\clearpage

\clearpage

\clearpage

\clearpage

\clearpage

\clearpage

\clearpage

\clearpage

\end{document}